\documentclass[a4 paper, 12 pt] {article}

\begin{document}
\title{\bf{Universality of the Linear Potential in Effective Models for the Low Energy QCD
coupled with the Dilaton Field}}
\author{Andrzej
Wereszczy\'{n}ski  \thanks{wereszcz@th.if.uj.edu.pl}
       \\
       \\ Institute of Physics,  Jagiellonian University,
       \\ Reymonta 4, 30-059 Krakow, Poland}
\maketitle
\begin{abstract}
QCD motivated effective models coupled with the cosmological
dilaton field are analyzed. It is shown that all models possess
confining solutions with the linear potential of confinement even
thought such solutions are not observed in the original effective
theory. In case of the Pagels-Tomboulis model analytical solutions
are explicit found.
\end{abstract}
%%%%%%%%%%%%%%%%%%%%%%%%%%%%%%%%%%%%%%%%%%%%%%%%%%%%%%%%%%%%%%%%%%%%%
\section{\bf{Introduction}}
%%%%%%%%%%%%%%%%%%%%%%%%%%%%%%%%%%%%%%%%%%%%%%%%%%%%%%%%%%%%%%%%%%%%%
The non-perturbative behavior of the quantum chromodynamics in the
low energy sector is one of the main problem of the temporary
theoretical physics. In spite of the many efforts such effects
like the confinement of quarks and gluons are not satisfactorily
understood. Because of the fact that we can not use the well known
perturbative methods we have to find a new way analyzing the
non-perturbative features. In fact, there are several methods of
investigating of the low energy QCD. Roughly speaking, they can be
divided into two groups. Namely, the lattice models and the
effective Ginzburg-Landau-like models. The second group contains
for example such popular models like the dual superconductor model
\cite{baker}, the color dielectric model \cite{lee} and the
stochastic vacuum model \cite{dosch}. In the present paper we will
focus on the second class of the effective models given by the
following ansatz for the lagrangian density in the Euclidean
space-time \cite{pagels}
\begin{equation}
L_{eff}= -\frac{1}{4} \frac{\mathcal{F}^{a \mu \nu }
\mathcal{F}^a_{\mu \nu }}{\bar{g}^2 (t)}, \label{effmodel}
\end{equation}
where
\begin{equation}
t=\ln \frac{\mathcal{F}}{\mu^4} \label{deft}
\end{equation}
and $\bar{g}$ is the running coupling constant and $\mu $ is a
dimensional constant. Here $\mathcal{F}=\frac{1}{2} F^a_{\mu \nu
}F^{a \mu \nu }$ and $F^a_{\mu \nu }=\partial_{\mu } A^a_{\nu }
-\partial_{\nu } A^a_{\mu } + g\epsilon^{abc} A^b_{\mu } A^c_{\nu
}$ is the standard field tensor depends on the $SU(2)$ gauge
fields. The model (\ref{effmodel}) was primary invented to
reproduce, at the classical level, the trace anomaly known from
the quantum chromodynamics \cite{pagels}. Indead, one can find
that trace of the energy-momentum tensor corresponding to
(\ref{effmodel}) reads
\begin{equation}
T^{\mu }_{\mu } = \frac{\beta (\bar{g})}{2 \bar{g}}
\frac{\mathcal{F}^{a \mu \nu } \mathcal{F}^a_{\mu \nu }}{\bar{g}^2
(t)}. \label{trace}
\end{equation}
This is in agreement with result obtained in the full quantum
theory. Particular examples of the Lagrangian (\ref{effmodel}) are
obtained by calculating $\bar{g}(t)$ from:
\begin{equation}
t=\int_{g_0}^{\bar{g}(t)} \frac{dg}{\beta (g)}. \label{beta}
\end{equation}
For example, the usual perturbative one-loop $\beta$-function
$\beta =-\frac{11}{32\pi } g^3$ gives the famous Savvidy-Adler
model \cite{savvidy}, \cite{adler}, \cite{mendel1}. For $\beta =
-\delta g$ one gets the Pagels-Tomboulis model \cite{pagels}. Both
models provide confinement of quarks and can be treated as first
approximation to the true effective model. Here confinement is
understood in the following way. Field configurations generated by
an external electric source possess infinite energy due to the
long range behavior of the fields. However, a dipole-like source
with zero total charge has finite energy. Because of that the
physical spectrum of the theory consists of the dipoles - mesons
whereas charge solutions are excluded from it.
\newline
The main aim of this paper is to analyze such effective models
together with the cosmological dilaton \cite{tree}. There are many
papers where classical solutions of the dilaton-Yang-Mills theory
has been considered \cite{lavre}, \cite{bizon}, \cite{dick1}.
However, because the classical Yang-Mills Lagrangian seems to have
little to do with physics described by the low energy QCD it is
probably more correct to investigate model where Yang-Mills part
is substituted by (\ref{effmodel}).
\newline
On the other hand, the non-minimal coupling between scalar and
gauge fields has been often used to reproduce the non-trivial
quantum phenomena in the framework of classical field theories
(c.f. color dielectric model). The scalar field represents unusual
properties of the non-perturbative vacuum in which the gauge
fields propagate. It can suggest that the correct effective model
should contain more than only gauge fields. It is in agreement
with lattice gauge theory where, in the low energy limit, many
additional effective fields (scalar, vector, tensor) appear
\cite{arodz}.
%%%%%%%%%%%%%%%%%%%%%%%%%%%%%%%%%%%%%%%%%%%%%%%%%%%%%%%%%%%%%%%%%%%%%
\section{\bf{The Linear Potential}}
%%%%%%%%%%%%%%%%%%%%%%%%%%%%%%%%%%%%%%%%%%%%%%%%%%%%%%%%%%%%%%%%%%%%%
Let us now couple an effective model for the low energy QCD with
the cosmological scalar field i.e. dilaton. Then the action, in
the Minkowski space-time, takes the following form:
\begin{equation}
L_{gen}= -\frac{e^{b\frac{\phi }{\Lambda } }}{4} \frac{ F^{a \mu
\nu } F^a_{\mu \nu }}{\bar{g}^2 (t)} +\frac{1}{2} \partial_{\mu }
\phi
\partial^{\mu } \phi ,\label{genmodel}
\end{equation}
where $t$ is given by $$t=\frac{1}{2} \ln \frac{F^2}{\Lambda^8}$$
and $b$ is a dimensionless constant. Here $\Lambda $ is a
dimensional constant and $F$ corresponds to $\mathcal{F}$ in the
Minkowski space-time. For convenience we introduce a new function
\begin{equation}
f(F):=\frac{1}{\bar{g}^2 (t)}. \label{def}
\end{equation}
The pertinent equations of motion read
\begin{equation}
D_{\mu } \left[ e^{b\frac{\phi }{\Lambda } } \frac{ \partial
(f(F)F)}{\partial F} F^{a \mu \nu } \right]= j^{a \nu }
\label{geqmot1}
\end{equation}
and
\begin{equation}
\partial_{\mu } \partial^{\mu } \phi = \frac{b}{2\Lambda } e^{b\frac{\phi}{\Lambda} }
f(F) F \label{geqmot2}
\end{equation}
In the present paper we are mainly interested in analyzing of the
field configurations generated by external electric static
sources. Due to that the external current is
\begin{equation}
j^{a \mu } =4\pi q \delta (r) \delta^{0\mu } \delta^{a3},
\label{source}
\end{equation}
where $q$ is an external charge. We would like to notice that
restriction to the Abelian current is not essential. Using the
results presented in \cite{dick1} one can find the solution of
these equations in the general non-Abelian case as well. However,
they differ from the Abelian solutions only by a multiplicative
color-dependent constant. The dependence on spatial coordinates
remains unchanged. One could argue that the small difference
between Abelian and non-Abelian case is in the contradiction with
the well-known fact that the non-perturbative features of QCD
originate in the non-Abelian character of the gauge fields. But
one should remember that there is no simple correspondence between
the quantum gauge fields in QCD and the fields in the presented
classical model. The non-Abelian character of the original quantum
theory is implemented by taking into account the QCD motivated
running coupling constant. The particular form of the Lie group on
which the classical gauge fields in the effective model are based
seems to play less important role (at least in the problem of the
confinement of external sources).
\\
Because of that we will consider only Abelian degrees of freedom.
For example one can set $A_{\mu }^a = A_{\mu } \delta^{a3}$. The
equations of motion can be rewritten as
\begin{equation}
\left[ r^2 e^{b\frac{\phi}{\Lambda } }  \frac{ \partial
(f(E^2)E^2)}{\partial E^2} E \right]'=4\pi q \delta(r)
\label{geqmot3}
\end{equation}
and
\begin{equation}
\nabla_r^2 \phi = -\frac{b}{2\Lambda } f(E^2) E^2
e^{b\frac{\phi}{\Lambda }}. \label{geqmot4}
\end{equation}
Here, we have assumed the spherical symmetry of the problem
$\vec{E}=E \hat{r}$. The prime denotes differentiation in respect
to $r$. The solutions take the following form
\begin{equation}
\frac{\phi (r)}{\Lambda } = -\frac{2}{b} \ln r\Lambda +\phi_0,
\label{gsol1}
\end{equation}
and
\begin{equation}
E=E_0 \Lambda^2 =const. \label{gsol2}
\end{equation}
Where $\phi_0$ and $E_0$ are yet to be determined. This
corresponds to linear electric potential
\begin{equation}
U=E_0r \Lambda^2.
\end{equation}
Inserting these solutions into the field equations we can obtain
the algebraic equations for the constants $\phi_0$ and $E_0$
\begin{equation}
e^{b\phi_0} \frac{ \partial f(E^2)E^2}{\partial E^2}
\left|_{E=E_0} E_0=q, \; \; \; e^{b\phi_0} f(E_0^2)
E^2_0=\frac{4}{b^2} \right.. \label{gcond1}
\end{equation}
After eliminating $\phi_0$ we find that the constant $E_0$ is
given by the following (in general non-linear) algebraic equation
\begin{equation}
E \frac{ \partial}{\partial E^2} \ln \left( f(E^2) E^2 \right)
\left|_{E=E_0} =\frac{qb^2}{4} \right. .\label{gcond2}
\end{equation}
Unfortunately we are not able to solve this equation in the
general case. However, as it is shown below, for particular forms
of the function $f$ one can find the value of constants $E_0$ and
$\phi_0$.
\newline
Let us analyze the energy of these solutions. The energy component
of the energy-momentum tensor corresponding to the Lagrangian
(\ref{effmodel}) has the following form
\begin{equation}
T_{00} = \left[ e^{b \frac{\phi }{\Lambda }} \left(E^2 \frac{
\partial (f(F) F)}{\partial F} -\frac{1}{4}f(F)F \right) +\frac{1}{2} (\partial_0
\phi )^2 + \frac{1}{2} (\partial_i \phi)^2\right].
\label{energyden}
\end{equation}
For the static, pure electric solutions we obtain
\begin{equation}
T_{00}= \left[ e^{b \frac{\phi }{\Lambda }} \left(E^2 \frac{
\partial (f(E^2) E^2)}{\partial E^2} -\frac{1}{2}f(E^2)E^2 \right) + \frac{1}{2} (\partial_i \phi)^2\right]. \label{energy1}
\end{equation}
Then after substituting the solutions (\ref{gsol1}), (\ref{gsol2})
and using relations (\ref{gcond1}) one gets
\begin{equation}
T_{00}=\frac{\Lambda^2}{r^2}\left[ e^{b \frac{\phi_0 }{\Lambda }}
\left(E_0^2 \frac{
\partial (f(E^2) E^2)}{\partial E^2} \left|_{E=E_0} \right. -\frac{1}{2} f(E_0^2) E_0^2 \right)
+ \frac{2}{b^2}\right]=\frac{\Lambda^2}{r^2} E_0 q.
\label{energy2}
\end{equation}
The energy stored in the ball with radius $R$ around the external
static, point-like electric charge diverges linearly with $R$
\begin{equation}
E(R)=\int_0^R T_{00} d^3r=4\pi E_0 q  R \Lambda^2. \label{energyR}
\end{equation}
The field configurations generated by external electric sources
have infinite total energy. In contradictory to the Maxwell theory
energy diverges due to the long range behavior of the fields. In
that sense electric charges are confined. It should be underlined
that appearance of the linear potential (constant energy density)
is independent on the form of the original effective Lagrangian.
One can find many examples of physically interesting models which
originally do not have linear potential. However, because of the
interaction with the dilaton field, the potential becomes linear.
%%%%%%%%%%%%%%%%%%%%%%%%%%%%%%%%%%%%%%%%%%%%%%%%%%%%%%%%%%%%%%%%%%%%%
\section{\bf{An Example}}
%%%%%%%%%%%%%%%%%%%%%%%%%%%%%%%%%%%%%%%%%%%%%%%%%%%%%%%%%%%%%%%%%%%%%
In order to find explicit solutions one has to choose a particular
form of the function $f$. Here we will take the Pagels-Tomboulis
model \cite{pagels}
\begin{equation}
L=-\frac{1}{4} \left( \frac{F^a_{\mu \nu } F^{a \mu \nu }
}{2\Lambda^4} \right)^{2\delta } F^a_{\mu \nu } F^{a \mu \nu },
\label{modelpt}
\end{equation}
where $\delta $ is a dimensionless parameter. It was shown that
the Pagels-Tomboulis model can serve as a candidate for the low
energy effective action for $\delta \in (\frac{1}{4}, \infty)$. In
particular, it assures confinement of electric charges and gives
the following confining potential \cite{my2}
\begin{equation}
U_{PT}= a_0 |q|^{\frac{2+4\delta}{1+4\delta}}
\Lambda^{\frac{8\delta}{1+4\delta}} r^{\frac{4\delta
-1}{4\delta+1}}, \label{potpt}
\end{equation}
which, after fitting the values of the parameter, is in agreement
with the experimental data. Here $a_0$ is a numerical constant.
One can easily see that the linear potential is achieved only in
the limit $\delta \rightarrow \infty $. Obviously such a limit can
not be implemented in the Lagrangian. That means that it is not
possible to realize the linear confinement in the Pagels-Tomboulis
model.
\newline
Let us turn to the model with the dilaton field. Then the
Lagrangian density reads
\begin{equation}
L=-\frac{e^{b\frac{\phi }{\Lambda }} }{4} \left( \frac{F^a_{\mu
\nu } F^{a \mu \nu } }{2\Lambda^4} \right)^{2\delta } F^a_{\mu \nu
} F^{a \mu \nu } +\frac{1}{2} \partial_{\mu } \phi \partial^{\mu }
\phi \label{model}
\end{equation}
Then equations of motion
\begin{equation}
D_{\mu } \left[ (2\delta +1) e^{b\frac{\phi }{\Lambda } }\left(
\frac{F}{\Lambda^4} \right)^{2\delta } F^{a \mu \nu } \right]=
j^{a \nu } \label{eqmot1}
\end{equation}
and
\begin{equation}
\partial_{\mu } \partial^{\mu } \phi = \frac{b}{2\Lambda } e^{b\frac{\phi}{\Lambda} }
\left( \frac{F}{\Lambda^4} \right)^{2\delta } F. \label{eqmot2}
\end{equation}
Similar as in the previous section we will consider an Abelian
static and point-like external electric charge. Thus the field
equations take the form
\begin{equation}
\left[ (2\delta +1)r^2 e^{b\frac{\phi}{\Lambda } }
\frac{E^{4\delta +1}}{\Lambda^{8\delta}} \right]'=q \delta(r)
\label{eqmot3}
\end{equation}
and
\begin{equation}
\nabla_r^2 \phi = -\frac{b}{2\Lambda } \frac{E^{4\delta
+2}}{\Lambda^{8\delta}} e^{b\frac{\phi}{\Lambda }}. \label{eqmot4}
\end{equation}
One can solve eq. (\ref{eqmot3}) and find the electric field
\begin{equation}
E=\Lambda^2\left( \frac{q}{(1+2\delta )r^2\Lambda^2 }
\right)^{\frac{1}{1+4\delta}} e^{-\frac{b}{1+4\delta }
\frac{\phi}{\Lambda } }.  \label{efield}
\end{equation}
Then eq. (\ref{eqmot4}) reads
\begin{equation}
\nabla_r^2 \phi=- \Lambda^3 \frac{b}{2} \left(
\frac{q^2}{(1+2\delta )^2 r^4 \Lambda^4 }
\right)^{\frac{1+2\delta}{1+4\delta}} e^{-\frac{b}{1+4\delta }
\frac{\phi }{\Lambda }}. \label{eqmot5}
\end{equation}
After some calculation one can find that the solution is
\begin{equation}
\frac{\phi (r)}{\Lambda } = -\frac{2}{b} \ln r\Lambda +\phi_0,
\label{sol1}
\end{equation}
where the constant
\begin{equation}
\phi_0 = -\frac{1+4\delta}{b} \ln \left[ \frac{4}{b^2}
\left(\frac{1+2\delta }{q} \right)^{2 \frac{1+2\delta
}{1+4\delta}} \right]. \label{stala}
\end{equation}
The corresponding electric field is given in the following form
\begin{equation}
E (r) = \frac{4(1+2\delta)}{qb^2} \Lambda^2. \label{sol2}
\end{equation}
Finally we obtain linear confining potential
\begin{equation}
U = \frac{4(1+2\delta)}{qb^2} r \Lambda^2. \label{sol3}
\end{equation}
As it was said before it is really remarkable that the potential
takes the linear form for all $\delta $. The dependence on the
parameter $\delta $ is, unlikely the original Pagels-Tomboulis
model, visible only in the 'string tension'. The functional
dependence is always linear.
\newline
One can notice that the case $\delta =0$ is a little bit special.
Then the additional symmetry appears in the equations of motion.
Namely, if one defines a new variable $x=\frac{1}{r}$ then the
translation $x \rightarrow x+x_0$ remains equations unchanged.
Because of that we find a whole family of the solutions which
depends on the translation parameter $x_0$ \cite{dick1}. These
solutions have finite energy. One should remember that this effect
is present only for $\delta =0$ and does not occur in the general
case.
%%%%%%%%%%%%%%%%%%%%%%%%%%%%%%%%%%%%%%%%%%%%%%%%%%%%%%%%%%%%%%%%%%%%%
\section{\bf{Conclusions}}
%%%%%%%%%%%%%%%%%%%%%%%%%%%%%%%%%%%%%%%%%%%%%%%%%%%%%%%%%%%%%%%%%%%%%
In the present paper we have shown that the cosmological dilaton
field coupled with the effective model (\ref{effmodel}),
originally dependent only on the $SU(2)$ gauge fields, provides
the linear confinement of the electric charges. It is striking
that this behavior is observed for all models based on gauge
fields with the $U(1)$ subgroup. Even though the original gauge
model does not possess confining solutions then interaction with
the dilaton causes that the electric charges are confined. The
confining potential is always linear and only string tension is
model dependent. In other words the particular form of the
effective model is not important if one would like to model the
confinement of the quarks. The essential is the form of the
coupling between the scalar field and the gauge field. The linear
confinement is not observed if one takes, for example power-like
coupling in stead of exponential one \cite{dick2}, \cite{my1} .
Then the potential strongly depends on the particular form of the
gauge part of the model. Only exponential coupling gives
indifferently the same functional form of the potential.
\newline
Of course, this result can be applied not only to QCD effective
models but to all model with at least $U(1)$ gauge field (for
instance the non-linear electrodynamics).
\newline
There are several directions in which the present work can be
continued. First of all one can analyze more general than the tree
level approximated dilaton coupling \cite{cvetic}. It would be
interesting to know how this more realistic interaction inflects
on the electric solutions of the QCD induced effective models.
Similar, one should consider the non-perturbative dilaton
potential \cite{nppot} and/or the mass term \cite{mass}. Moreover,
one could also ask about another cosmological scalar field i.e.
the modulus field \cite{cvetic}, \cite{modulus}. However, in our
opinion the most important problem is to analyze dipole sources.
As we have mentioned it before, disappearance of the electric
charge from physical spectrum is not sufficient to have
confinement in the theory. One has to show that fields generated
by a dipole-like source possess finite energy. It would be
interesting to know whether the functional form of the total
energy is also in this case independent on the gauge part of the
model. We plan to address this last problem in out next paper.
\\
\\
We would like to thank Professor H. Arod\'{z} for reading the
manuscript and many helpful remarks.


\begin{thebibliography}{30}
\bibitem{baker} M. Baker, J. S. Ball, F. Zachariasen, Phys. Rep.
{\bf 209}, 73 (1991); M. Baker, N. Brambilla, H. G. Dosch, A.
Vairo, Phys. Rev. D {\bf 58}, 034010 (1998).
\bibitem{lee} R. Friedberg, T. D. Lee, Phys. Rev. D {\bf 15}, 1694
(1977); D {\bf 18}, 2623 (1978).
\bibitem{dosch} H. G. Dosch, Y. A. Simonov, Phys. Lett. B {\bf
205}, 339 (1988); Y. A. Simonov, Nucl. Phys. B {\bf 307}, 512
(1988).
\bibitem{pagels} H. Pagels, E. Tomboulis, Nucl. Phys. B {\bf 143}, 485 (1978).
\bibitem{savvidy} G. Matinyan, G. K. Savvidy, Nucl. Phys. B {\bf
134}, 539 (1978).
\bibitem{adler} S. L. Adler, Phys. Rev. D {\bf 23}, 2905 (1981);
S. L. Adler, Phys. Lett. B {\bf 110}, 302 (1981); S. L. Adler, T.
Piran, Phys. Lett. B {\bf 113}, 405 (1982); S. L. Adler, T. Piran,
Phys. Lett. B {\bf 117}, 91 (1982).
\bibitem{mendel1} P. M. Fishbane, S. Gasiorowicz, P. Kaus, Phys.
Rev. D {\bf 36}, 251 (1987); D {\bf 43}, 933 (1991); R. R. Mendel
et al., Phys. Rev. D {\bf 30}, 621 (1984); D {\bf 33}, 2666
(1986); D {\bf 40}, 3708 (1989); D {\bf 42}, 911 (1990).
\bibitem{tree} G. W. Gibbons, K. Maeda, Nucl. Phys. B {\bf 298},
741 (1988); D. Garfinkle, G. T. Horowitz, A. Strominger, Phys.
Rev. D {\bf 43}, 3140 (1991); G. T. Horowitz, "The dark side of
string theory: black holes and black strings", in {\it Proceedings
of the 1992 Trieste Spring School on String Theory and Quantum
Gravity}.
\bibitem{lavre} G. Lavrelashvili, D. Maison, Phys. Lett. B {\bf
295}, 67 (1992).
\bibitem{bizon} P. Bizon, Phys. Rev. D {\bf 47}, 1656 (1993).
\bibitem{dick1} R.Dick, Phys. Lett. B {\bf 397}, 193 (1996); R.Dick Phys. Lett B {\bf 409}, 321 (1997).
\bibitem{arodz} H. B. Nielsen, A. Patcos, Nucl. Phys. {\bf B 195}, 137 (1982);
H.-J. Pirner, J. Wroldsen M. Ilgenfritz, Nucl. Phys. {\bf B 294},
905 (1987); J.-F. Mathiot, G. Chanfray, H.-J. Pirner, Nucl. Phys.
{\bf A 500}, 605 (1989); H. Arodz, H.-J. Pirner, Acta Phys. Pol.
{\bf B 30}, 3895 (1999).
\bibitem{my2} H. Arod\'{z}, M. \'{S}lusarczyk, A. Wereszczy\'{n}ski, Acta Phys.
Pol. B {\bf 32}, 2155 (2001).
\bibitem{dick2} R. Dick, L. P. Fulcher, Eur. Phys. J. C
{\bf 9}, 271 (1999); R. Dick Eur. Phys. J. C {\bf 6}, 701 (1999).
\bibitem{my1} M. \'{S}lusarczyk, A. Wereszczy\'{n}ski, hep-ph/0307213, Eur. Phys. J. C, in press;
Eur. Phys. J. C {\bf 28}, 151
(2003); Eur. Phys. J. C {\bf 23}, 145 (2002); Acta Phys. Pol. B
{\bf 32}, 2911 (2001).
\bibitem{cvetic} M. Cvetic, A. A. Tseytlin, Nucl. Phys. B {\bf 416}, 137
(1994).
\bibitem{nppot} J. Garcia-Bellido, M. Quir\'{o}s, Nucl. Phys. B
{\bf 385}, 558 (1992); N. R. Stewart, Mod. Phys. Lett. A {\bf 7},
983 (1992); R. Brustein, P. Steinhardt, Phys. Lett. B {\bf 302},
196 (1993).
\bibitem{mass} J. H. Horne, G. T. Horowitz, Nucl. Phys. B {\bf
399}, 169 (1993).
\bibitem{modulus} E. Witten, Phys. Lett. B {\bf 155}, 151 (1985).
\end{thebibliography}
\end{document}